\documentclass[12pt]{article}
\usepackage{amssymb} 
\usepackage{amsmath} 
\usepackage{amsthm}
\usepackage{amsfonts}
\usepackage{graphicx}

\begin{document}
\font\frak=eufm10 scaled\magstep1
\font\fak=eufm10 scaled\magstep2
\font\fk=eufm10 scaled\magstep3
\font\black=msbm10 scaled\magstep1
\font\bigblack=msbm10 scaled\magstep 2
\font\bbigblack=msbm10 scaled\magstep3
\font\scriptfrak=eufm10
\font\tenfrak=eufm10
\font\tenblack=msbm10


\def\biggoth #1{\hbox{{\fak #1}}}
\def\bbiggoth #1{\hbox{{\fk #1}}}
\def\sp #1{{{\cal #1}}}
\def\goth #1{\hbox{{\frak #1}}}
\def\scriptgoth #1{\hbox{{\scriptfrak #1}}}
\def\smallgoth #1{\hbox{{\tenfrak #1}}}
\def\smallfield #1{\hbox{{\tenblack #1}}}
\def\field #1{\hbox{{\black #1}}}
\def\bigfield #1{\hbox{{\bigblack #1}}}
\def\bbigfield #1{\hbox{{\bbigblack #1}}}
\def\Bbb #1{\hbox{{\black #1}}}
\def\v #1{\vert #1\vert}             
\def\ord#1{\vert #1\vert} 
\def\m #1 #2{(-1)^{{\v #1} {\v #2}}} 
\def\lie #1{{\sp L_{\!#1}}}               
\def\pd#1#2{\frac{\partial#1}{\partial#2}}
\def\pois#1#2{\{#1,#2\}}
\def\set#1{\{\,#1\,\}}             
\def\<#1>{\langle#1\rangle}        
\def\>#1{{\bf #1}}                
\def\f(#1,#2){\frac{#1}{#2}}
\def\cociente #1#2{\frac{#1}{#2}}
\def\braket#1#2{\langle#1\mathbin\vert#2\rangle} 
\def\brakt#1#2{\langle#1\mathbin,#2\rangle}           
\def\dd#1{\frac{\partial}{\partial#1}} 
\def\bra #1{{\langle #1 |}}
\def\ket #1{{| #1 \rangle }}
\def\ddt#1{\frac{d #1}{dt}}
\def\dt2#1{\frac{d^2 #1}{dt^2}}
\def\matriz#1#2{\left( \begin{array}{#1} #2 \end{array}\right) }
\def\Eq#1{{\begin{equation} #1 \end{equation}}}

\def\bw{{\bigwedge}}      
\def\hut{{\scriptstyle \land}}            
\def\dg{{\goth g^*}}                                                                                                            
\def\Cdg{{C^\infty (\goth g^*)}}
\def\poi{\{\:,\}}                           
\def\qw{\hat\omega}                
\def\FL{{\sp F}L}                 
\def\hFL{\widehat{{\sp F}L}}      
\def\XHMw{\goth X_H(M,\omega)} 
\def\XLHMw{\goth X_{LH}(M,\omega)}                  
\def\ea{\varepsilon_a}
\def\ep{\varepsilon}
\def\mitad{\frac{1}{2}}
\def\x{\times}  
\def\cinf{C^\infty} 
\def\forms{\bigwedge}                 
\def\onda{\tilde}
\def\orb{{\sp O}}

\def\a{\alpha}
\def\d{\delta}
\def\g{{\gamma }}                  
\def\G{{\Gamma}}	
\def\La{\Lambda}                   
\def\la{\lambda}                   
\def\w{\omega}                     
\def\W{{\Omega}}                   
\def\ltimes{\bowtie}

\def\roc{{\tilde{\cal R}}}                       
\def\cl{{\cal L}}                               
\def\V{{\sp V}}                                 
\def\F{{\sp F}}
\def\cv{{{\goth X}}}                    
\def\LG{\goth g}
\def\LH{\goth h}
\def\X{{{\goth X}}}                     
\def\R{{\hbox{{\field R}}}}             
\def\big R{{\hbox{{\bigfield R}}}}
\def\bbig R{{\hbox{{\bbigfield R}}}}
\def\C{{\hbox{{\field C}}}}         
\def\Z{{\hbox{{\field Z}}}}             
\def\N{{\hbox{{\field N}}}}         

\def\ima{\hbox{{\rm Im}}}                               
\def\dim{\hbox{{\rm dim}}}        
\def\End{\hbox{{\rm End}}} 
\def\Tr{\hbox{{\rm Tr}}} 
\def\tr{{\hbox{\rm\small{Tr}}}}                
\def\lin{{\hbox{Lin}}}
\def\vol{{\hbox{vol}}}  
\def\Hom{{\hbox{Hom}}}
\def\rank{{\hbox{rank}}}
\def\Ad{{\hbox{Ad}}}
\def\ad{{\hbox{ad}}}
\def\CoAd{{\hbox{CoAd}}}
\def\coad{{\hbox{coad}}}                           
\def\Rea{\hbox{Re}}                     
\def\id{{\hbox{id}}}                    
\def\Id{{\hbox{Id}}}
\def\Int{{\hbox{Int}}}
\def\Ext{{\hbox{Ext}}}
\def\Aut{{\hbox{Aut}}}
\def\Card{{\hbox{Card}}}
\def\SODE{{\small{SODE }}}
\newcommand{\bea}{\begin{eqnarray}}
\newcommand{\eea}{\end{eqnarray}}

\def\R{\mathbb{R}}
\def\ba{\begin{eqnarray}}
\def\ea{\end{eqnarray}}
\def\be{\begin{equation}}
\def\ee{\end{equation}}
\def\Eq#1{{\begin{equation} #1 \end{equation}}}
\def\R{\Bbb R}
\def\C{\Bbb C}
\def\Z{\Bbb Z}
\def\a{\alpha}                  
\def\b{\beta}                   
\def\g{\gamma}                  
\def\d{\delta}                  
\def\bra#1{\langle#1|}
\def\ket#1{|#1\rangle}
\def\goth #1{\hbox{{\frak #1}}}
\def\<#1>{\langle#1\rangle}
\def\cotg{\mathop{\rm cotg}\nolimits}
\def\Map{\mathop{\rm Map}\nolimits}
\def\wt{\widetilde}
\def\const{\hbox{const}}
\def\grad{\mathop{\rm grad}\nolimits}
\def\Div{\mathop{\rm div}\nolimits}
\def\braket#1#2{\langle#1|#2\rangle}
\def\Erf{\mathop{\rm Erf}\nolimits}
\def\matriz#1#2{\left( \begin{array}{#1} #2 \end{array}\right) }
\def\Eq#1{{\begin{equation} #1 \end{equation}}}
\def\deter#1#2{\left| \begin{array}{#1} #2 \end{array}\right| }
\def\pd#1#2{\frac{\partial#1}{\partial#2}}
\def\til{\tilde}

\def\la#1{\lambda_{#1}}
\def\teet#1#2{\theta [\eta _{#1}] (#2)}
\def\tede#1{\theta [\delta](#1)}
\def\N{{\frak N}}
\def\GR{{\cal G}}
\def\Wei{\wp}

\def\frac#1#2{{#1\over#2}} \def\pd#1#2{\frac{\partial#1}{\partial#2}}
\def\matrdos#1#2#3#4{\left(\begin{matrix}#1 & #2 \cr          
                                 #3 & #4 \cr\end{matrix}\right)}

\newtheorem{teor}{Teorema}[section]
\newtheorem{cor}{Corolario}[section]
\newtheorem{prop}{Proposici\'on}[section]
\newtheorem{note}[prop]{Note}
\newtheorem{definicion}{Definici\'on}[section]
\newtheorem{lema}{Lema}[section]
\theoremstyle{plain}
\newtheorem{theorem}{Theorem}
\newtheorem{corollary}{Corollary}
\newtheorem{proposition}{Proposition}
\newtheorem{definition}{Definition}
\newtheorem{lemma}{Lemma}

\def\Eq#1{{\begin{equation} #1 \end{equation}}}
\def\R{\Bbb R}
\def\C{\Bbb C}
\def\Z{\Bbb Z}
\def\mp#1{\marginpar{#1}}

\def\la#1{\lambda_{#1}}
\def\teet#1#2{\theta [\eta _{#1}] (#2)}
\def\tede#1{\theta [\delta](#1)}
\def\N{{\frak N}}
\def\Wei{\wp}
\def\Hil{{\cal H}}

\font\frak=eufm10 scaled\magstep1

\def\bra#1{\langle#1|}
\def\ket#1{|#1\rangle}
\def\goth #1{\hbox{{\frak #1}}}
\def\<#1>{\langle#1\rangle}
\def\cotg{\mathop{\rm cotg}\nolimits}
\def\cotanh{\mathop{\rm cotanh}\nolimits}
\def\arctanh{\mathop{\rm arctanh}\nolimits}
\def\wt{\widetilde}
\def\const{\hbox{const}}
\def\grad{\mathop{\rm grad}\nolimits}
\def\Div{\mathop{\rm div}\nolimits}
\def\braket#1#2{\langle#1|#2\rangle}
\def\Erf{\mathop{\rm Erf}\nolimits}

\centerline{\Large \bf Applications of Lie systems in dissipative}
\bigskip
\centerline{\Large \bf  Milne--Pinney equations.}
\vskip 0.75cm

\centerline{ Jos\'e F. Cari\~nena and Javier de Lucas}
\vskip 0.5cm

\centerline{Departamento de  F\'{\i}sica Te\'orica, Universidad de Zaragoza,}
\medskip
\centerline{50009 Zaragoza, Spain.}
\medskip

\vskip 1cm

\begin{abstract}
We use the geometric approach to the theory of Lie systems of differential
equations  in order to study dissipative Ermakov systems. We prove 
 that there is a superposition rule for solutions of such equations. This fact
 enables us  
 to express the general solution of a dissipative Milne--Pinney equation in
 terms of particular solutions of a system of second-order linear differential
 equations and a set of constants.
\end{abstract}

\section{Introduction}

\qquad An instance of nonlinear equations that has been receiving an increasing  interest
during the last years because of its ubiquity in physics and engineering 
 is the presently  called 
Milne--Pinney equation  \cite{{Mil30},P50},
\begin{equation}\label{MP}
\ddot x=-\omega^2(t)x+\frac k{x^3}\,,
\end{equation}
where $k$ is a real constant with values depending up on the field in which the
equation is applied.
Ermakov introduced this equation when looking for a first integral for the corresponding time-dependent
harmonic oscillator \cite{Er80}. In a very short paper, \cite{P50}, Pinney showed that the
 general solution of (\ref{MP}) can be written in terms of a pair of solutions 
of the corresponding harmonic oscillator and two constants. More recently it has been shown that the differential equation (\ref{MP}) admits a
superposition rule involving any two particular solutions and two constants \cite{CL08}.

Lately Haas \cite{Ha07} started studying how to find
an approximate solution for the simplest damped Pinney equation, the one with a
  damping term linear  in the velocity. Moreover other problems with a term
  with a quadratic
dependence on the  velocity have recently been studied \cite{Mu08a,{MuRoy},SQBG}.

The geometric theory of Lie systems \cite{LS}-\cite{CGM07} has been very efficient in dealing
with  equation (\ref{MP}) and the corresponding Ermakov system \cite{Er80,CL08, CLR08}. The possibility
of considering a larger class of systems that can be reduced to a Lie system 
has been shown recently. As a particular example we can study
 a
generalization of the Pinney equation in which the constant $k$ is replaced by a time-dependent
function. So we can deal with an equation similar to (\ref{MP})
but including a term proportional to the velocity because such an equation can
be reduced to one like (\ref{MP}) in which  the constant $k$ is replaced by a
time-dependent term. The latter suggests revisiting the theory of Lie systems from this 
new perspective in order to deal with such dissipative Milne--Pinney
equations. This is the aim of the paper, which is organized as follows:
We study in Section 2 the possibility of removing the term linear in velocity by means of a time-dependent
change of coordinates  while the way of doing a similar
thing by means of a time-reparametrization is analyzed in Section 3. 
The properties of Lie systems are used in Section 4 to establish the
 corresponding
nonlinear superposition rules for some time-dependent genealizations of 
Milne--Pinney equations. Finally in Section 5 the recently proposed theory of quasi-Lie systems 
is applied in the study  of differential equations of  such type.
\section{Damped harmonic oscillator with time-dependent angular frequency}

\qquad In this Section we show that we can remove terms proportional to velocity in a
certain sort of second-order differential equation through a simple
time-dependent transformation. As a main result we deal with a dissipative
Milne--Pinney equation in order to remove its dissipative term in this
way. Nevertheless this time-dependent change of variables is an {\it ad-hoc}
method and we just explain here how it works. Later we will explain that
quasi-Lie schemes explain this transformation \cite{CGL08}. We also obtain this
result and more information about some dissipative Milne--Pinney equations through the theory of quasi-Lie systems  without this {\it ad-hoc} 
assumption.

As a first example we consider an harmonic oscillator with a time-dependent angular frequency,
$\omega(t)$, and a term proportional to the velocity with coefficient $\gamma(t)$, namely
\begin{equation}
\ddot x+\gamma(t)\,\dot x+\omega^2(t) \, x=0\,.\label{tdtdho}
\end{equation}
Particular instances of this equation are the ones with a damping term with constant coefficient
$\gamma_0$,
\begin{equation}
\ddot x+\gamma_0\,\dot x+\omega^2(t) \, x=0\,,\label{dtdho}
\end{equation}
or the interesting example, found in \cite{Pedrosa}, given by the choice 
$$\gamma(t)=-\frac{d}{dt}\log [f(t)]\,,
$$
which describes a physical model with variable mass $m(t)=[f(t)]^{-1}$.

Consider a time-dependent transformation of the variable $x$ of the form 
\begin{equation}
x= \zeta(t)\,y\label{scchange}
\end{equation}
which leads to 
$$
\dot x=  \zeta(t)\,\dot y+ \dot  \zeta(t)\,y\,,
\qquad 
\ddot x= \zeta(t)\,\ddot y+2\,\dot  \zeta(t)\,\dot y+\ddot  \zeta(t)\,y\,. 
$$
This time-dependent change of variables transforms the equation (\ref{tdtdho})
into
\begin{equation}\label{eq24}
\ddot y+\left(\gamma(t)+2\,\frac{\dot \zeta(t)}{\zeta(t)}\right)\,\dot y+
\left(\frac{\ddot \zeta(t)}{\zeta(t)}+\gamma(t)\frac{\dot \zeta(t)}{\zeta(t)}+\omega^2(t)
\right) \,y=0\,.
\end{equation}

Therefore we can eliminate the term proportional to velocity by using a
function $\zeta(t)$ such that
$$\gamma(t)+2\,\frac{\dot \zeta(t)}{\zeta(t)}=0
$$
or, more explicitly,
\begin{equation}
\zeta(t)=\zeta_0 \exp\left(-\frac 12\int_0^t\gamma(t')\ dt'\right)\,.\label{elimvel}
\end{equation}
In this case equation (\ref{tdtdho}) reduces to the harmonic oscillator 
\begin{equation}
\ddot y+\Omega^2(t)\ y=0\,,\label{tdhoeq}
\end{equation}
with an
angular frequency $\Omega(t)$ given by 
\begin{equation}
\Omega^2(t)=\omega^2(t)-\frac{\gamma(t)^2}{4}-\frac{\dot\gamma(t)}2
\label{Omegat}
\end{equation}
because, in view of $\dot \zeta(t) =-\frac 12  \gamma(t)\zeta(t)$, we obtain
$$\frac{\ddot \zeta(t)}{\zeta(t)}
=-\frac{\dot\gamma(t)}2-\frac{\gamma(t)}2\left(-\frac{\gamma(t)}2\right)
=\frac{\gamma^2(t)}4-\frac{\dot\gamma(t)}2
\,,
$$
and then
$$
\frac{\ddot \zeta(t)}{\zeta(t)}+\gamma(t)\frac{\dot \zeta(t)}{\zeta(t)}=-\frac{\gamma^2(t)}4-\frac{\dot\gamma(t)}2\,.
$$
Hence equation (\ref{tdtdho}) is transformed into
 the time-dependent  harmonic oscillator (\ref{tdhoeq}) 
with the time-dependent angular frequency
$\Omega(t)$ given by (\ref{Omegat}). Note that equation (\ref{tdhoeq}) is the reduced canonical form of (\ref{tdtdho}) as it is 
indicated in \cite{Milson,Berko}.

We have found a time-dependent transformation which enables one to remove the term proportional to velocity in (\ref{tdtdho}). We study some particular instances of this method. For instance, if $\gamma(t)=\gamma_0$, the
transformation 
\begin{equation}
x= e^{-\frac{\gamma_0}2 t}\,y\,,\label{constantLiou}
\end{equation}
transforms the differential equation (\ref{tdtdho}) into
\begin{equation}
\ddot y+\left(\omega^2(t)-\frac{\gamma_0^2} 4\right)y=0\,.\label{tdepho}
\end{equation}
So we can analyze the damped system (\ref{dtdho}) through the time-dependent harmonic oscillator (\ref{tdhoeq}) with a time-dependent angular frequency
$\Omega(t)$ given by 
$$
\Omega(t)=\omega^2(t)-\frac{\gamma_0^2} 4\,.
$$
This latter nonautonomous second-order differential equation has been
considered quite often both in the classical and in the quantum approach (see
e.g. \cite{BL, CLR08c,{CV}}) and we can deal with it by means of the theory of Lie
systems \cite{CLR08,CLR08c,{CLR07a}}.

Had we started with a generalized Pinney equation with a
 time-dependent
coupling $k(t)$,
\begin{equation}
\ddot x+\gamma(t)\,\dot x+\omega^2(t) \, x=\frac {k(t)} {x^3}\,,\label{gdMPeq}
\end{equation}
we would obtain, through the same time-dependent change of variables (\ref{scchange}), the following equation 
$$
\ddot y+\left(\gamma(t)+2\,\frac{\dot \zeta(t)}{\zeta(t)}\right)\,\dot y+
\left(\frac{\ddot \zeta(t)}{\zeta(t)}+\gamma(t)\frac{\dot \zeta(t)}{\zeta(t)}+\omega^2(t)
\right) \,y=\frac{k(t)}{\zeta^4(t)\, y^3}\,.
$$
Moreover, if $\zeta(t)$ is defined by (\ref{elimvel}), the preceding equation
simplifies to the velocity-independent differential equation
\begin{equation}
\ddot y+\Omega^2(t)\ y=\frac{k(t)}{\zeta^4(t)\, y^3}\,.\label{tdMP}
\end{equation}
Here $\Omega(t)$ is given by (\ref{Omegat}) and $\zeta$ is the function (\ref{elimvel}).

For the usually called damped
Milne--Pinney equation, $\gamma(t)=\gamma_0$, the transformation (\ref{constantLiou})
  reduces the differential equation (\ref{dtdho}) to  
$$
\ddot y+\left(\omega^2(t)-\frac{\gamma_0^2} 4\right)y=\frac{k(t)e^{2\gamma_0 t}}{y^3}\,.
$$

Notice that in the particular case of equation (\ref{gdMPeq}) with $\gamma
(t)=\gamma_0$ and $k(t)=k_0\,e^{-2\gamma_0 t}$, we recover
a Milne--Pinney equation like (\ref{MP}) with $k_0$ and $\Omega(t)$ instead of
$k$ and 
$\omega(t)$. This is exactly the example considered in \cite{Na86} as
associated with the Caldirola--Kanai model.

 In the more general case of $\gamma(t)$ being an arbitrary function, if $\zeta (t) $
 is chosen to be such that the term proportional to the velocity vanishes,
i.e. $k(t)=k_0\,\exp(-2\int^t\gamma(t')\,dt')$, we also recover 
a Milne--Pinney equation like (\ref{MP}) with coefficient $k_0$.

To sum, for a generic function $k(t)$ we can remove the term proportional to
the  velocity,
but the coefficient of the nonlinear term becomes time-dependent.

\section{Time-reparametrization of some second-order differential equations.}

\qquad In the preceding  section we studied a particular time-dependent 
change of variables which allows us to eliminate the term proportional to
velocities in a certain kind of second-order differential equation. Now we
show
 that this can also be done by means of a time-reparametrization.

Given the second-order differential equation,
$$\ddot x=f(x,\dot x,t),$$
the time-reparametrization given by a new parameter $s$ such that 
\begin{equation}
\frac{ds}{dt}=\alpha(t),\label{sdet}
\end{equation}
where $\alpha(t)$ has a constant sign, for instance  $\alpha(t)$ is  positive,
which defines a good reparametrization allowing us to express $s$ as a function
of $t$ and, conversely,  $t$ as a function
of $s$,  
 leads to 
$$x'=\frac{dx}{ds}=\frac 1{\alpha(t)}\, \dot x\ \Longleftrightarrow \ \dot
x=\alpha(t)\, x'
$$
and therefore
$$
\ddot x=\dot \alpha(t)\,x'+\alpha^2(t)\, x''\,.
$$
So we get 
$$x''=\frac 1{\alpha^2(t(s))}\left(f(x,\alpha(t(s))\, x',t(s))-\dot \alpha(t(s)) \,x'\right)\,.
$$

In the particular instance of a second-order differential equation of the type
$$\ddot x=a(t)\,\dot x+b(t)\, x+\frac{c(t)}{x^3}
$$
the transformed equation is
$$
x''=\frac 1{\alpha^2(t(s))}\left(a(t(s))\,\dot x+b(t(s))\,
  x+\frac{c(t(s))}{x^3}\right)-\frac{\dot\alpha(t(s))}{\alpha^2(t(s))}\, x'\,,
$$
i.e.
$$x''=\frac 1{\alpha(t(s))}\left(a(t(s))-\frac {\dot
    \alpha(t(s))}{\alpha(t(s))}\right)\, x'+
\frac{b(t(s))}{\alpha^2(t(s))}\,x+
\frac {c(t(s))}{\alpha^2(t(s))\, x^3}\,.
$$

Note that, if the function $\alpha(t)$ is chosen to be given by
$$
\alpha(t)=\alpha_0\exp\left(\int_0^t a(t')\ dt'\right),
$$
the term containing the new velocity disappears and an equation of
Milne--Pinney type
 is obtained. Nevertheless the coefficient of the nonlinear term is
$s$-dependent
$$
x''=\frac{b(t(s))}{\alpha^2(t(s))}\,x+
\frac {c(t(s))}{\alpha^2(t(s))\, x^3}\,.
$$
So, we have obtained a reduction process from the dissipative Milne--Pinney equation into 
 an equation of
Milne--Pinney type in a new way.

\section{Lie systems of second-order differential equations.}

\qquad We remark that a system of second-order differential equations 
$$\ddot x^i=f^i(x,\dot x,t)\,,\qquad i=1,\ldots,n,
$$
is associated with a system of first-order differential equations 
with a double number of variables, namely 
\begin{equation}
\left\{
\begin{array}{rcl}
\dot x^i&=&v^i,\\
\dot v^i&=&f^i(x,v,t),
\end{array} \qquad i=1,\ldots,n.
\right.
\end{equation}
Even more generally the system of $n$ second-order differential equations 
corresponding to 
the system of $2n$ first-order differential equations 
\begin{equation}
\left\{
\begin{array}{rcl}
  \dot x^i&=&\alpha(t) v^i,\\
\dot v^i&=&f^i(x,v,t),
\end{array} \qquad i=1,\ldots,n,
\right.
\end{equation}
is related to
$$\ddot x^i=\frac{\dot \alpha}{\alpha}\dot x^i+\alpha\, f^i(x,\dot
x/\alpha,t), \qquad i=1,\ldots,n\,.
$$
This type of equations may appear as Schr\"odinger equations of
position-dependent mass
\cite{AtkSev} with the von Roos prescription \cite{vR}.

As particular instances,  the damped time-dependent harmonic oscillator described by
(\ref{dtdho}) is associated with the system of first-order differential equations
\begin{equation}
\left\{
\begin{array}{rcl}
  \dot x&=&e^{-\gamma_0 t} v,\\
\dot v&=&-e^{\gamma_0 t}\omega^2(t) \, x\,,
\end{array} 
\right.
\end{equation}
and the damped generalized Milne--Pinney equation (\ref{gdMPeq}) 
is related to the system
\begin{equation}\left\{
\begin{array}{rcl}
  \dot x&=&e^{-\gamma_0 t} v\\
\dot v&=&e^{\gamma_0 t}\left(-\omega^2(t) \, x+\dfrac {k(t)} {x^3}\right)
\end{array}
\right.\,.
\end{equation}

Recall that the standard Milne--Pinney equation (\ref{MP}) 
 is linked to
 the following system of
first-order differential equations
\begin{equation}\label{FOMP}
\left\{\begin{aligned}
\dot x&=v,\\
\dot v&=-\omega^2(t)x+\frac{k}{x^3}.
\end{aligned}\right. 
\end{equation}
Hence we can study the second-order differential equation (\ref{MP})
through the system
 (\ref{FOMP}). The solutions of 
(\ref{FOMP}) are integral curves for the time-dependent vector field
\begin{equation}
X(t)=v\frac{\partial}{\partial x}+\left(-\omega^2(t)x+\frac{k}{x^3}\right)\frac{\partial}{\partial v},
\end{equation}
which can be written as a linear combination 
\begin{equation}\label{vf}
X(t)=X_2-\omega^2(t)X_1.
\end{equation}
of the vector fields
\begin{equation}
X_1=x\frac{\partial}{\partial v},\quad X_2=v\frac{\partial}{\partial
  x}+\frac{k}{x^3}\frac{\partial}{\partial v}.
\end{equation}
These vector fields close on a real Lie algebra isomorphic to $\goth{sl}(2,\mathbb{R})$
with the vector field 
\begin{equation}
 X_3=\frac{1}{2}\left(x\frac{\partial}{\partial x}-v\frac{\partial}{\partial v}\right)\,,
\end{equation}
because the vector fields $X_\alpha$  satisfy the
commutation relations
\[
[X_1,X_2]=2X_3,\quad [X_3,X_2]=-X_2,\quad [X_3,X_1]=X_1\,.
\]
Therefore the second-order differential equation (\ref{MP}) is a SODE Lie
system in the sense of 
\cite{CLR08,CLR07a} and we can use the theory of Lie systems in order to study its properties.

More generally each time-dependent vector field of the form
\begin{equation}
X(t)=\beta (t)\, X_1+\alpha(t)\,X_2\,,
\end{equation}
where $\alpha$ and $\beta$ are arbitrary time-dependent functions, is a Lie system
of the same type
 as the one studied in \cite{CLR08, CLR07a}. Its integral curves 
are the solutions of the system
\begin{equation}
\left\{\begin{aligned}
\dot x&=\alpha(t)\, v,\\
\dot v&=\beta(t)\,x+\alpha(t)\,\frac{k}{x^3}.
\end{aligned}\right. \label{Fsystem}
\end{equation}
This system of differential equations is a generalization of the case given in \cite{CLR08,CLR07a} and
 is related to the  second-order differential equation,
\begin{equation}\label{set}
\ddot x-\frac{\dot \alpha(t)}{\alpha(t)}\dot x-\alpha(t)\beta(t)\,x-\alpha^2(t)\frac{k}{x^3}=0,
\end{equation}
because the derivative of the first equation is 
$$\ddot x=\dot \alpha(t)\,v+ \alpha(t)\dot v,$$
and, in view of (\ref{Fsystem}), we obtain (\ref{set})

If $\alpha$ is positive and we use the notations
$\alpha=e^{-F}$ and $\beta=-qe^F$,  equation
 (\ref{set}) becomes the more general differential equation studied in
\cite{Re99},
\begin{equation}
\ddot x+\dot F\, \dot x+q\, x-\frac 1{e^{2F}}\,\frac k{x^3}=0.
 \end{equation}

The latter equation can be studied by means of
the theory
 of Lie systems and a  superposition rule for its general solution can be
 found. 
In order to get such a rule we have to consider two copies of the  Lie system,
\begin{equation}\label{FOHO}
\left\{\begin{aligned}
\dot x&=e^{-F}v\\
\dot v&=-q\,e^{F}\,x
\end{aligned}\right. ,
\end{equation}
which corresponds to the second-order differential equation
\begin{equation}
\ddot x+\dot F\dot x+qx=0,
\end{equation}
together with  one copy of (\ref{FOMP}). Thus we obtain the system of first-order differential equations
\[
\left\{
\begin{array}{rcl}
\dot x&=&e^{-F}v_x,\cr
\dot y&=&e^{-F}v_y,\cr
\dot z&=&e^{-F}v_z,\cr
\dot v_x&=&-qe^{F}\,x+e^{-F}\dfrac{k}{{x^3}},\cr
\dot v_y&=&-qe^{F}\,y,\cr
\dot v_z&=&-qe^{F}\,z ,
\end{array}\right.
\]
which corresponds to the differential equations of the integral curves for the time-dependent vector field 
\[X=e^{-F}\left(v_x\frac{\partial}{\partial x}+v_y\frac{\partial}{\partial y}+v_z\frac{\partial}{\partial z}+\frac{k}{x^3}
\frac{\partial}{\partial v_x}\right)-      qe^{F}\,       \left(x\frac{\partial }{\partial v_x}+y\frac{\partial }{\partial v_y}+
z\frac{\partial}{\partial v_z}\right)\,.
\]
This vectorfield  can be expressed as $X=e^{-F}N_2-{qe^{F}}N_1$, 
where $N_1$ and $N_2$ are
\[
N_1=y\frac{\partial }{\partial v_y}+x\frac{\partial }{\partial v_x}+
z\frac{\partial}{\partial v_z},\quad
 N_2=v_x\frac{\partial}{\partial x}+v_y\frac{\partial}{\partial y}+v_z\frac{\partial}{\partial z}+
\frac{1}{x^3}\frac{\partial}{\partial v_x}.
\] 
These vector fields generate a 3-dimensional real Lie algebra with the
vector field $N_3$ given by 
\[
 N_3=\frac 12\left(x\frac{\partial}{\partial x}+y\frac{\partial}{\partial y}
+z\frac{\partial}{\partial z}
-v_x\frac{\partial}{\partial
     v_x}
-v_y\frac{\partial}{\partial   v_y}
-v_z\frac{\partial}{\partial v_z}\right)\,.
\]
In fact, they generate a Lie algebra isomorphic to $\mathfrak{sl}(2,\mathbb{R})$ because  
 \[
[N_1,N_2]=2N_3, \quad [N_3,N_1]=N_1, \quad  [N_2,N_3]=N_2\,.
\]

The dimension of the distribution generated by these vector fields is three 
and the manifold of the Lie system has dimension six. There are three
 time-independent integrals of motion which turn out to be
the Ermakov invariant
 $I_1$ of the subsystem involving variables $x$ and $y$, 
the Ermakov invariant $I_2$ 
of the subsystem involving variables $x$ and $z$, and  
the Wronskian $W$ of the subsystem involving variables $y$ and $z$.
They define a  foliation  with 3-dimensional leaves. 
 This foliation  can be used to obtain a superposition rule.

The Ermakov invariants are
\[
I_1=\frac 12\left((yv_{x}-xv_y)^2+k\left(\frac {y}x\right)^2\right)\,,\qquad
I_2=\frac 12\left((xv_{z}-zv_x)^2+k\left(\frac {z}x\right)^2\right)\,,
\]
where $I_1$ and $I_2$ are non-negative constants for $k>0$ and the Wronskian $W$ is
\[
W=yv_{z}-zv_{y}\,.
\]

It is to be remarked that the relation of the first equation (\ref{FOHO})
allows us to rewrite the invariants $I_1$ and $I_2$, respectively, as:
$$
I_1=\frac 12\!\left(e^{2F(t)}(y\dot x-x\dot y)^2+k\left(\frac {y}x\right)^2\right),\quad
I_2=\frac 12\!\left(e^{2F(t)}(x\dot z-z\dot x)^2+k\left(\frac {z}x\right)^2\right).
$$

In particular, for $F(t)=\gamma_0\,t$, which corresponds to the
above mentioned example  associated with the Caldirola--Kanai
model,
we recover the invariant given in (2.11b) of \cite{Na86}. 

 We can obtain an explicit expression of $x$ in terms of $y, z$ and the
 three first integrals $I_1, I_2, W$
\[
x=\frac {\sqrt 2}{\mid W\mid}\left(I_2y^2+I_1z^2\pm\sqrt{4I_1I_2-kW^2}\ yz\right)^{1/2}\,.
\]	

We remark that $W$ is a constant fixed by the two independent particular
solutions of the time-dependent harmonic oscillator $x_1(t)$ and $x_2(t)$, i.e.
$y(t)=x_1(t)$ and $z(t)=x_2(t)$, and
only $I_1$ and $I_2$ 
play the role of constants in this  superposition rule
for the Milne--Pinney equation. This is not a surprising fact
 because the Milne--Pinney equation is a second-order differential equation. Note also that the values of $I_1$ and $I_2$ are  nonnegative constants,
but should be chosen such that  $x(0)$ be real, i.e. $4I_1I_2\geq kW^2$.

\section{Ermakov systems that are not of Lie system type}

\qquad In this section we show that there exist dissipative Milne--Pinney equations that can be transformed into simple Milne--Pinney equations by means 
of the methods developed for quasi-Lie systems. Furthermore we can deal with the so-obtained 
Ermakov systems as in \cite{CLR07a} in order to find 
many of their properties, i.e. first integrals of motion or
 superposition rules. Next we can use these results to obtain properties of the initial system of
 differential equations by inverting the time-dependent change
 of variable.

Consider the family of differential equations
\begin{equation}\label{eq1}
\ddot x=a(t)\dot x+b(t) x+c(t)\frac{1}{x^3}\,.
\end{equation}

We are mainly interested in the case $c(t)\not =0$ and we can assume  that
$c(t)$
has a constant sign for the set of values of $t$ we are considering. The case
in which $c(t)$ is identically zero corresponds to the  harmonic
oscillator with time-dependent frequency and a dissipative term.

Usually we associate with such a second-order  differential 
equation a  system of first-order differential equations by introducing a new
variable,  $v$, and relating (\ref{eq1}) to the system of first-order
differential equations 
\begin{equation}\label{eq2}\left\{
\begin{array}{rcl}
\dot x&=&v,\\
\dot v&=&a(t)v+b(t) x+c(t)\dfrac{1}{x^3}\,.\\
\end{array}\right.
\end{equation}
This system describes the integral curves for the time-dependent vector field 
$$X(t)=a(t)X_1+b(t)X_2+c(t)X_3+X_4\,,$$
where 
the  vector fields $X_1,\ldots,X_4$, are given by
\begin{equation}\label{VectBasis}
X_1=v\pd{}{v},\quad X_2=x\pd{}{v},\quad X_3=\frac{1}{x^3}\pd{}{v},\quad
X_4=v\pd{}{x}\,.
\end{equation}
Consider also the vector field 
\begin{equation}
 X_5=x\pd{}{x}\,.
\end{equation}
The set of these  five vector fields is a basis for a $\R$-linear space $V$. However,
they do not close on a Lie algebra because the commutator
$\left[X_3,X_4\right]$ is not in $V$. 
Moreover, it can be checked that there is no finite-dimensional real
Lie algebra $V'$ containing $V$. Therefore the differential equation (\ref{eq2}) cannot be considered
 as a Lie system. Nevertheless we can deal with this differential equation through a quasi-Lie scheme. 

The two-dimensional linear subspace, $W\subset V$,   generated by 
the  vector fields
\begin{equation}
Y_1=X_1=v\pd{}{v},\qquad Y_2=X_2=x\pd{}{v}\,,
\end{equation}
is a Lie algebra because 
these vector fields satisfy the commutation relation
\begin{equation}
\left[Y_1,Y_2\right]=-Y_2\,.
\end{equation}
What is more, as 
\begin{equation}
\begin{array}{lll}
\left[Y_1,X_3\right]=-X_3,&\quad \left[Y_1,X_4\right]=X_4,&\quad\left[Y_1,X_5\right]=0\,,\cr
\left[Y_2,X_3\right]=0,&\quad \left[Y_2,X_4\right]=X_5-X_1,&\quad\left[Y_2,X_5\right]=-X_2,
\end{array}
\end{equation}
the linear space $V$ is invariant under the action of  Lie algebra $W$ on $V$,
i.e. 
 $\left[W,V\right]\subset V$. Therefore we have found a quasi-Lie scheme to deal with the differential equation (\ref{eq2}).

The corresponding set of time-dependent diffeomorphisms of ${\rm T}\R$ related to the
flows
 of time-dependent vector fields in $W$,
$\alpha_1(t)Y_1+\alpha_2(t)Y_2$,
is given by
\begin{equation}\left\{
\begin{array}{rcl}
x&=&x',\\
v&=&\alpha(t)v'+\beta(t)x'
\end{array}\right.\label{transf}
\end{equation}
with $\alpha(t)\neq 0$. The inverse transformation is 
\begin{equation}\left\{
\begin{array}{rcl}
x'&=&x,\\
v'&=&-\dfrac{\beta(t)}{\alpha(t)}x+\dfrac 1{\alpha(t)}v.
\end{array}\right.\label{invtransf}
\end{equation}

These time-dependent diffeomorphisms transform
 the system (\ref{eq2}) into a new one in which the time-dependent vector field
 determining the dynamics can be written as a linear combination of the fields 
of $V$  at each time
\begin{equation}\label{fineq}
X'(t)=a'(t)X_1+b'(t)X_2+c'(t)X_3+d'(t)X_4+e'(t)X_5\,.
\end{equation}
More explicitly the new coefficients are 
\begin{equation}
\left\{\begin{aligned}
a'(t)&=a(t)-\beta(t)-\frac{\dot \alpha(t)}{\alpha(t)},\\
b'(t)&=\frac{b(t)}{\alpha(t)}+a(t)\frac{\beta(t)}{\alpha(t)}-\frac{\beta^2(t)}
{\alpha(t)}-\frac{\dot\beta(t)}{\alpha(t)},\\
c'(t)&=\frac{c(t)}{\alpha(t)},\\
d'(t)&=\alpha(t),\\
e'(t)&=\beta(t),\\
\end{aligned}\right.
\end{equation}
and the integral curves for (\ref{fineq}) are solutions of the system
\begin{equation}\label{quasiErmsys}\left\{
\begin{array}{rcl}
\dfrac{dx'}{dt}&=&\beta(t)x'+\alpha(t)v',\cr
\dfrac{dv'}{dt}&=&\dfrac{\beta(t)}{\alpha(t)}\left(\dfrac{b(t)}{\beta(t)}+a(t)-\beta(t)
-\dfrac{\dot\beta(t)}{\beta(t)}\right) x'+\left(a(t)-\beta(t)-\dfrac{\dot\alpha(t)}{\alpha(t)}\right)v'\cr&+&\dfrac{c(t)}{\alpha(t)}\dfrac{1}{x'^3}.
\end{array}\right.
\end{equation}

However, notice that only, if $\beta(t)=0$, is this system associated with a
second-order differential equation, more specifically with
\begin{equation*}
\frac{d^2 x'}{dt^2}=a(t)\,\frac{dx'}{dt}+b(t)\,x'+c(t)\frac{1}{x'^3}\,.
\end{equation*}

The Ermakov systems studied in \cite{CLR07a} are in the family of
differential equations (\ref{eq2}). Hence it is natural to look for sufficient
conditions to
be able to  transform a given system of (\ref{eq2}) into one of these Ermakov systems
of the form
\begin{equation}\label{Ermsys}\left\{
\begin{array}{rcl}
\dot x&=&f(t)v,\\
\dot v&=&-\omega^2(t) x+f(t)\dfrac{k}{x^3},\,
\end{array}\right.
\end{equation}
 where $k$ is a constant, which corresponds to the second-order differential equation
\begin{equation}\label{Erm}
\ddot x=\frac{\dot f(t)}{f(t)}\,\dot x+f(t)\,\left(-\omega^2(t) x+f(t)\dfrac{k}{x^3}\right)  \,.
\end{equation}
Next we compare (\ref{quasiErmsys}) with (\ref{Ermsys}) to transform 
equation (\ref{eq2}) into one related to Ermakov systems. As a result we
notice that $\alpha=f$ and the time-dependent coefficients $a(t)$ and $c(t)$
must be such that  
\begin{equation}\label{alfaeq}\left\{
\begin{array}{rcl}
k\alpha(t)&=&\dfrac{c(t)}{\alpha(t)},\\
\dfrac{\dot\alpha(t)}{\alpha(t)}&=&a(t),
\end{array}\right.
\end{equation}
i.e. the sign of $k$ must coincide with that of $c(t)$ and 
$$\omega^2(t)=-\sqrt{\dfrac{k}{c(t)}}b(t).$$

This expression provides a sufficient condition in order to be able to transform a
differential equation corresponding to the
 system (\ref{eq2}) into
 one of the form of (\ref{Erm}).  Taking the time-derivative on the first condition of
 (\ref{alfaeq}) we obtain $2\,k\,\alpha\, \dot\alpha=\dot c$. Dividing by
 $\alpha^2$ and using the second condition in (\ref{alfaeq}) we get 
\begin{equation}\label{IC}
a(t)=\dfrac{\dot c(t)}{2 c(t)}.
\end{equation}
Thus the resulting transformation is determined by
\begin{equation}
\alpha(t)=\sqrt{\frac{c(t)}{k}},\qquad \beta(t)=0,
\end{equation}
for a certain constant $k$.

Next we point out some  differential equations
(\ref{eq1}) appearing in the literature that can be related by means of the method
developed here with  a particular Lie system: the Milne--Pinney equation   \cite{CLR07a}.

As a first example we analyze the Chini differential equation \cite{Ch98}
\begin{equation}\label{Chini}
\ddot x+\frac{\dot p(t)}{2p(t)}\dot x+\frac{q(t)}{p(t)}x=\frac
1{p(t)}\frac{k}{x^3},\qquad {\rm  with}\quad p(t)>0\,.
\end{equation} 
This equation is associated with the system of first-order differential equations
\begin{equation}\label{Chinisys}\left\{
\begin{aligned}
\dot x&=v,\\
\dot v&=-\frac{\dot p(t)}{2 p(t)}v-\frac{q(t)}{p(t)}x+\frac{1}{p(t)}\frac{k}{x^3}\,.
\end{aligned}\right.
\end{equation}
This system is a particular instance of  
(\ref{eq2}) for the following choice of time-dependent coefficients
\begin{equation}
\begin{aligned}
a(t)=-\frac{\dot p(t)}{2p(t)},\quad  b(t)=-\frac{q(t)}{p(t)} , \quad
c(t)=\frac{k}{p(t)}\ .
\end{aligned}
\end{equation}
In this case, as $\dot c/(2c)=-\dot p/(2p)=a$, these coefficients satisfy the reducibility condition
(\ref{IC}) and
we can transform this system into a Lie one through the transformation (\ref{transf}) determined by the coefficients 
\begin{equation}
\alpha(t)=\dfrac{1}{\sqrt{p(t)}}\,,  \qquad
\beta(t)=0\,,
\end{equation}
i.e. by means of  the time-dependent change of variables
\begin{equation}\left\{
\begin {aligned}
x&=x',\\
v&=\dfrac{1}{\sqrt{p(t)}}v'.\,
\end{aligned}\right.
\end{equation}
So equation (\ref{Chinisys}) becomes
\begin{equation}\label{LieChini}\left\{
\begin{array}{rl}
\dfrac{dx'}{dt}&=\dfrac{1}{\sqrt{p(t)}}v',\\
\dfrac{dv'}{dt}&=-\dfrac{q(t)}{\sqrt{p(t)}}x'+\dfrac{1}{\sqrt{p(t)}}\dfrac{k}{x'^3}.
\end{array}\right.
\end{equation}
This system describes the integral curves for the time-dependent vector field
$$
X(t)=\frac 1{\sqrt{p(t)}}\left[v'\,\pd{}{x'}+\left(-q(t)\, x'+\dfrac{k}{x'^3}\right)\,\pd{}{v'} \right]\,.
$$ 

Introducing the time-reparametrization
\begin{equation}
\tau(t)=\int^t_0\frac{dt'}{\sqrt{p(t')}}
\end{equation}
the system (\ref{LieChini}) reduces to the Ermakov system  studied in \cite{CLR07a}
\begin{equation}\label{ErmChini}\left\{
\begin{array}{rl}
\dfrac{dx'}{d\tau}&=v',\\
\dfrac{dv'}{d\tau}&=-{q(t(\tau))}x'+\dfrac{k}{x'^3}\,.
\end{array}\right.
\end{equation}
 which corresponds to the second-order differential equation
\begin{equation}
\frac{d^2x'}{d\tau^2}=-q(t(\tau))x'+\dfrac{k}{x'^3}\,.
\end{equation}

Now we can use the Ermakov invariant, superposition rules etc. for
(\ref{ErmChini}). Next, by inverting the time-dependent change of variables
used for the Chini equation, we obtain some kind of time-dependent
superposition rule for the solutions of (\ref{Chini}). 

Walter \cite{Wa68} developed another interesting example: 
\begin{equation}\label{walter}
\ddot x+\frac{\dot p(t)}{p(t)}\dot x+\frac{q(t)}{p(t)}x=\frac{k}{p^2(t) x^3}.
\end{equation} 
This equation is associated with the system of first-order differential equations
\begin{equation}\label{WalterSys}\left\{
\begin{aligned}
\dot x&=v,\\
\dot v&=-\frac{\dot p(t)}{p(t)}v-\frac{q(t)}{p(t)}x+\frac{1}{p^2(t)}\frac{k}{x^3}\,.
\end{aligned}\right.
\end{equation}
This equation is a particular instance of (\ref{eq2}) with the functions
\begin{equation}
a(t)=-\frac{\dot p(t)}{p(t)},\quad  b(t)=-\frac{q(t)}{p(t)} , \quad c(t)=\frac{k}{p^2(t)}\,.
\end{equation}
In this case, as \,$\dot c/c=-2\dot p/p$,  these functions satisfy the
reducibility 
  condition (\ref{IC}) and 
 if we choose
\begin{equation}
\alpha(t)=\frac{1}{p(t)}\,,\qquad
\beta(t)=0\,,
\end{equation}
then the transformation (\ref{transf}) transforms 
the equation (\ref{WalterSys}) into the Lie system
\begin{equation}\label{LieWalter}\left\{
\begin{array}{rcl}
\dfrac{dx'}{dt}&=&\dfrac{1}{p(t)}v',\\
\dfrac{dv'}{dt}&=&-q(t)x'+\dfrac{1}{p(t)}\dfrac{k} {x^3}
\,.
\end{array}\right.
\end{equation}

Introducing a new time function $\tau$ by means of 
\begin{equation}
\tau(t)=\int^t_0\frac{dt'}{p(t')}
\end{equation}
we obtain
\begin{equation}
\frac{d^2x}{d\tau^2}=-q(t(\tau))p(t(\tau))x'+\frac{k}{x'^3},
\end{equation}
which is a standard Milne--Pinney  equation 
 that can be described again through the theory of such systems as in \cite{CLR07a}. 

Finally we consider the example of Colegrave and Abdalla \cite{CA83}:
\begin{equation}
\ddot x-2 \frac{\dot p(t)}{p(t)}\dot x+p^2(t)x=p^4(t)\frac{k}{x^3},
\end{equation} 
with an associated system of first-order differential equations
\begin{equation}
\left\{
\begin{aligned}
\dot x&=v,\\
\dot v&=\frac{2\dot p(t)}{p(t)}v-p^2(t)x+p^4(t)\frac{k}{x^3}\,,
\end{aligned}\right.
\end{equation}
which is a particular instance of (\ref{eq1}) with the coefficients
\begin{equation}
\begin{aligned}
a(t)=\frac{2\dot p(t)}{p(t)},\quad  b(t)=-p^2(t) , \quad c(t)=kp^4(t)\,.
\end{aligned}
\end{equation}
In this   case, if we take $\beta(t)=0$ and we choose as $\alpha$ a  solution of the differential equation,
\begin{equation}
\frac{\dot\alpha(t)}{\alpha(t)}-a(t)=0,
\end{equation}
as
\begin{equation}
\alpha(t)=p^2(t)\,,
\end{equation}
we obtain that (\ref{transf}) transforms the initial differential equation  into
\begin{equation}\label{eq9}\left\{
\begin{aligned}
\frac{dx'}{dt}&=p^2(t)v',\\
\frac{dv'}{dt}&=-x'+p^2(t)\frac{k} {x^3}\,,
\end{aligned}\right.
\end{equation}
which can be studied through the formalism of Lie systems. On the introduction
of  a new
time function 
\begin{equation}
\tau(t)=\int^t p^2(t')\,dt'
\end{equation}
the last differential equation  reduces to
\begin{equation}
\frac{d^2x'}{d\tau^2}=-\frac{1}{p^2(t(\tau))}x'+\frac{k}{x'^3}
\end{equation}
and the usual Milne--Pinney equation is recovered.
\section{Conclusions and Outlook}

\qquad We have studied ' damped time-dependent angular frequency harmonic oscillators'
and time-reparametrizations
 of second-order differential equations through {\it ad hoc}
 transformations. As a result we have 
found a method  to remove terms proportional to the velocity in certain set of
second-order differential equations. Afterwards we  revisited the theory
of SODE Lie systems and quasi-Lie schemes. These theories allow us to study
many
(systems of)  second-order differential equations. They provide methods to
obtain
 time-dependent superposition rules, integrals of motion, solutions etc. In this paper we have applied these theories to some particular dissipative Ermakov systems to recover known properties from a new point of view.

The theory of quasi-Lie schemes can be used to deal with many systems of differential equations. We expect to analyse new differential equations in forthcoming papers, recovering known properties and finding new ones.

\section*{Acknowledgements}
The authors are grateful to Profs. J.M.F. Bassalo, F. Haas and P.G.L. Leach  for critical lecture of the manuscript and valuable comments. 
 Partial financial support by research projects MTM2006-10531 and E24/1 (DGA) 
 are acknowledged. JdL also acknowledge
 a F.P.U. grant from the  Ministerio de Educaci\'on y Ciencia.

\end{document}